\documentclass[11pt]{article}
\usepackage{mathpazo}

\usepackage{amsmath,bm}
\usepackage{graphicx,pdflscape}
\usepackage{color}
\usepackage[toc,page]{appendix}
\usepackage{simpler-wick}

\usepackage{color,subfigure}

\usepackage{hyperref}
\hypersetup{
    colorlinks,
    citecolor=red,
    filecolor=cyan,
    linkcolor=blue,
   urlcolor=magenta
 }

\newcommand{\disp}[1]{Eq.~(\ref{#1})}

\newcommand{\figdisp}[1]{Fig.~(\ref{#1})}

\newcommand{\lessim} {\ {\raise-.5ex\hbox{$\buildrel<\over\sim$}}\ }

\newcommand{\gssim}{\ {\raise-.5ex\hbox{$\buildrel>\over\sim$}}\ }

\newcommand{\si}{\sigma}
\newcommand{\sib}{\bar{\sigma}}
\newcommand{\tJ}{\ $t$-$J$ \ }

\renewcommand{\emph}{\textit}

\usepackage{hyperref}
\usepackage{color}

\newcommand{\change}[1]{{\color{black} #1}}

\newcommand{\beq}{\begin{eqnarray}}
\newcommand{\eeq}{\end{eqnarray}}
\newcommand{\barray}{\begin{eqnarray}}
\newcommand{\earray}{\end{eqnarray}}

\begin{document}

\title{ Aspects of the Normal State Resistivity of  Cuprate Superconductors}

\author{ B Sriram Shastry$^{1,}$\footnote{sriram@physics.ucsc.edu} \; and Peizhi Mai$^{2,1}$ \\
\small \em $^{1}$Physics Department, University of California, Santa Cruz, CA, 95064 \\
\small \em $^{2}$CNMS, Oak Ridge National Laboratory,  Oak Ridge, TN, 37831-6494
}
\date{ February 3, 2020}
\maketitle

\abstract{   Planar normal state  resistivity data  taken from three families of cuprate superconductors  are compared with theoretical calculations from the recent extremely correlated Fermi liquid theory (ECFL) \cite{ECFL}. The two hole doped cuprate materials $LSCO$ and $BSLCO$ and the electron doped material $LCCO$ have yielded rich data sets  at several densities $\delta$ and temperatures T, thereby  enabling a systematic comparison with theory. The  recent ECFL resistivity calculations for the highly correlated  $t$-$t'$-$J$ model by us give the resistivity  for a wide set of model parameters \cite{SM,MS}.
After using  X-ray diffraction and angle resolved photoemission  data to fix parameters appearing in the theoretical resistivity, only one  parameter, the magnitude of the hopping $t$, remains undetermined. For each  data set,  the slope of the experimental resistivity at a {\em single} temperature-density  point  is sufficient to determine $t$, and hence the resistivity on absolute scale at  all  remaining densities and temperatures. This procedure is    shown to give a fair account of the entire  data.
}

\newpage

{\bf \S Introduction:}

\vspace{.15 in}

Understanding the normal state resistivity of  high-T$_c$ cuprate superconductors and other   strongly correlated materials  is a challenging problem. The resistivity reveals the nature of the lowest energy  charge excitations and therefore constitutes a relatively simple  and yet  fundamental probe of matter.  In cuprates the different   chemical compositions, conditions of preparation, temperatures and a wide range of electronic densities  leads to a complex  variety of  data sets. These are  almost impossible to understand within the standard Fermi liquid theory  of metals. Major puzzles are the almost T-linear planar  resistivity of the hole-doped cuprates, the T$^2$ resistivity of the closely related electron-doped cuprates and  the intermediate behavior at various densities.  Indeed one of the larger questions about the cuprates is whether  the differing T dependence of the electron-doped and hole-doped  cases can possibly arise from a common physical model.  Equally  puzzling is the drastic reduction of the observed T scale of the resistivity variation ($\sim$100-400 K) from a bare bandwidth ($\sim$eV's) by a few orders of magnitude for both electron-doped and hole-doped cuprates. This situation has generated an upsurge of often radically new theoretical  work on correlated systems in the last three  decades, amounting to something like a  revolution in  condensed matter physics.
  In these  new class of theories 
the planar resistivity stands at the center \cite{ECFL,SM,MS,Th-1,Th-2,Th-3,Th-4,Th-5,Th-6,Th-7,Th-8,Th-9,Th-10,Th-11},  its unusual temperature dependence is most often  emphasized.

In this work  we  bring   theory face to face with  experimental data  on resistivity. We focus on
the  extremely correlated Fermi liquid theory (ECFL)  proposed by Shastry\cite{ECFL,ECFL-2, ECFL-expts}, where a  detailed and meaningful  comparison has become possible, as explained below.  Starting from a microscopic Hamiltonian, the ECFL theory yields the resistivity on an absolute scale with a very few parameters determining the underlying model. 
The resistivity is calculated  starting from   the $t$-$t'$-$J$ model\cite{tJ,PWA,Ogata} containing  four  parameters, of which three parameters  can be fixed using  ARPES  and X-ray  crystal structure  data, thus only {\it one } parameter remains undetermined. The theory  works in 2-dimensions without introducing { any}  redundant degrees of freedom, and therefore  the results can be meaningfully  tested against data on a variety of cuprates, including both hole-doped and electron-doped cases.

\vspace{.15 in}

{\bf \S Summary of the ECFL theory:}

\change{
 A  summary of the basic ideas and context of the ECFL theory  is provided here, readers  familiar with these ideas may skip to the later   sections giving the results.  The ECFL formalism  is applicable in any dimension  to doped Mott-Hubbard systems described by the $t$-$t'$-$J$ model \cite{tJ,PWA,Ogata}
 \beq
 H= - t \!\!\sum_{<i,j>} (\widetilde{C}^\dagger_{i \sigma} \widetilde{C}_{j \sigma}+h.c.) - t' \!\!\! \sum_{<\!<i,j>\!>} (\widetilde{C}^\dagger_{i \sigma} \widetilde{C}_{j \sigma}+h.c.) + J\!\! \sum_{<i,j>} \left( \vec{S}_i.\vec{S}_j - \frac{1}{4} n_i n_j\right), \label{H}
 \eeq
 where $<\!i,j\!> (<\!<\!i,j\!>\!>)$ denotes a sum over nearest (next-nearest) neighbors $i,j$,  the Gutzwiller projector is given by  $P_G= \prod_i(1- n_{i \uparrow} n_{i \downarrow})$, the operator  $\widetilde{C}_{i \sigma}= P_G {C}_{i \sigma} P_G $ is the Gutzwiller projected version of the standard (canonical) Fermion operator and $\vec{S}_i$ ($n_i$) the spin (density) operator at site i. 
  This model is in essence obtained from the Hubbard model by a  canonical transformation  implementing   the large U limit \cite{tJ}. The transformation preserves the physics of the strong coupling  Hubbard model at the lowest energies.   The large energy scale $U$ of the Hubbard model is traded for non-canonical anticommutation relations between Gutzwiller projected electrons in the $t$-$t'$-$J$ model.
  
 Standard (Feynman) diagrammatic many-body techniques do not apply  to  the \tJ model due to  the effect of the Gutzwiller projection on the anticommutation relations.  For the relevant operators $\widetilde{C}, \widetilde{C}^\dagger$ of the  model \disp{H}, the canonical Fermionic anticommutator $\{C_{i \si_i},C^\dagger_{j \si_j}\}= \delta_{ij}\delta_{\si_i\si_j}$ is replaced by a non-canonical anticommuting  Lie algebra
\beq 
\{ \widetilde{C}_{i \sigma_i},  \widetilde{C}^\dagger_{j \sigma_j} \}= \delta_{ij} \left( \delta_{\sigma_i \sigma_j}- \sigma_i \sigma_j  \widetilde{C}^\dagger_{j \sib_i}  \widetilde{C}_{i \sib_j}\right), \label{non-can}
\eeq
  where $\sib_i= - \si_i$. An immediate resulting problem  is that Wick's theorem  simplifying products of operators into pairwise contractions is now invalid. Hence a  formally exact  and  systematic Feynman-Dyson series expansion of the Greens functions in a suitable parameter is unavailable. On the other hand, in the Hubbard model with canonical Fermions, the Feynman-Dyson series exists but  is not controllable since $U$ the parameter of expansion is very  large for  strong correlations.  In trading  the Hubbard model for the \tJ model in the Gutzwiller projected subspace,  we gain the   tactical advantage of  avoiding  accounting for the large energy scale. However this advantage is lost  unless we succeed in finding  a corresponding formally exact expansion to replace the Feynman-Dyson series.  
  
  The ECFL formalism solves this problem by replacing the Feynman-Dyson series with an alternate $\lambda$ series. This  series is    formally exact and is  an expansion of the Greens functions 
   in a parameter $\lambda$. This parameter lives in a finite domain $\lambda \in[0,1]$, interpolating between the free Fermi gas at $\lambda=0$  and the fully Gutzwiller projected limiting case $\lambda=1$. One way is to  introduce $\lambda$ as the coefficient the  non-canonical term  in the anti-commutator \disp{non-can}. For analogy  it is useful to compare \disp{non-can} with  the contrast between the commutators of canonical Bosons and   usual rotation group ($SU(2)$) Lie algebra  of spin-S particles. One finds \cite{S-pathintegral}  that  $\lambda$ plays a  parallel role to the inverse spin, in the theory of  quantum spin systems, i.e.  $\lambda \leftrightarrow \frac{1}{2 S}$ where $S=\frac{1}{2},1,\ldots$.
  
For computing the Greens functions, we note the  {\em exact} functional differential equation of the canonical Hubbard model and the \tJ model written in shorthand space-time-spin matrix notation\cite{ECFL,SM} as
\beq
\left(g^{-1}_0 - U  \frac{\delta}{\delta {\cal V}} - U G \right).G&=&\!\! \delta \mathbb{1}, \label{Hub} \\
\left(g^{-1}_0- \lambda \hat{X} - \lambda \hat{Y}_1  \right).G&=&\!\! \delta (\mathbb{1}- \lambda \gamma),\;\; \label{tJH}
\eeq   
where $g_0^{-1}$ is the non-interacting Greens function, $\gamma$ is a local version of $G$, and the remaining terms (of a similar character as the 2nd and 3rd term in \disp{Hub}) are detailed in \cite{ECFL,SM}. Here \disp{Hub} is the functional differential equation for the Hubbard model. By inverting the operator multiplying $G$ and expanding in U, one  generates the complete Feynman series in powers of $U$  for the Hubbard model.  In \disp{tJH}  $\lambda$ is set at unity to obtain the exact equation for the \tJ model. Its  iteration of the above type is not straightforward due to the extra time dependent term on the right hand side. 
 These are the equations of motion in the presence of a space-time-spin dependent potential ${\cal V}$, which is set at zero at the end as prescribed in the Schwinger-Tomonaga method of field theory. The Fermionic antiperiodic boundary conditions on $G$  in the  imaginary time variable complete the mathematical statement of the problem.

The ECFL formalism converts non-canonical equations \disp{tJH} into a pair of equations of the type \disp{Hub} by introducing a decomposition of the Greens function  $G= g . \widetilde{\mu}$ into  auxiliary Greens function $g$ and a caparison function $\widetilde{\mu}$. These pieces satisfy
the exact equations
\beq
\left(g^{-1}_0- \lambda {\wick{ \c X . \c g } }.\;  g^{-1} - \lambda \hat{Y}_1  \right). g& =&\!\! \delta \mathbb{1}\label{geq} \\
\widetilde{\mu} = \delta \  ( \mathbb{1} - \lambda {\gamma}  ) + \lambda \wick{\c X. g. \c {\widetilde{\mu}  }} \label{mueq}
\eeq
where the contraction symbol indicates that the functional derivative contained in $X$ acts on the term at the other end of the symbol, while other terms satisfy matrix  product rules.
Notice that the \disp{geq} looks similar to \disp{Hub} with a unit matrix on the right hand side, and is thus essentially  like a canonical Greens function expression. The second equation \disp{mueq} must be solved simultaneously with \disp{geq}, since $\hat{Y}_1$ depends on both $g$ and $\widetilde{\mu}$.  This task is done by expanding all variables systematically in powers of $\lambda$ and writing down a set of successive equations to each order. The solution thus found is continuously connected to the free Fermi gas, and  satisfies the Luttinger Ward volume theorem at $T=0$.  The latter is an essential part of claiming  that the resulting theory is a variety of Fermi liquid, being notoriously difficult to satisfy in uncontrolled approximations such as the truncations of Greens function equations.

 On setting the time dependent potential to zero we get the frequency dependent Greens function as
\beq 
G(k, i \omega_j)=
  g(k, i \omega_j) \times \widetilde{\mu}(k, i \omega_j)=  \frac{1- \lambda \frac{n}{2} + \lambda \Psi(\vec{k}, i\omega_j)}{g_{0}^{(-1)}(\vec{k}, i\omega_j)  - \lambda \Phi(\vec{k}, i\omega_j)}  \label{eq1} 
\eeq
where the two self energies $\Psi,\Phi$  determine G.  The ECFL formalism has  a systematic  expansion of these equations in powers of $\lambda$, starting with the free Fermi gas as the lowest term and finally setting $\lambda=1$.

 An expansion in $\lambda$ thus
  provides  a controlled framework for explicit calculations \cite{ECFL,ECFL-2}.
   The current version of the theory \cite{ECFL,ECFL-2,SM,MS} is valid to ${\cal O}(\lambda^2)$ and has been benchmarked against other standard techniques for strong coupling in limiting cases of infinite  dimensionality (i.e. DMFT)  and the single impurity limit \cite{ECFL-2}. Higher order terms in $\lambda$  are expected to impact the results outside  the regime considered here, namely $0.13 \lessim \delta \lessim 0.2$. It has been recently applied to  several   objects of experimental interest  such as angle resolved photoemission  (ARPES), Raman scattering, optical conductivity and the Hall constant and  recently the resistivity \cite{ECFL-expts,SM,MS}. }

\change{One of the main effects of strong correlations is to reduce significantly the quasiparticle weight $Z$ from its Fermi gas value of unity.  It is  worth commenting that the exact dynamical mean field theory (DMFT) studies of the Hubbard model in $d=\infty$  using a  mapping to a self consistent Anderson impurity model yield a very small $Z$  for $U> 2.918\, D$ (2D is the bandwidth) as one approaches  the insulating limit $n \to 1$. This is seen e.g. in Fig.~(1.a) of \cite{ECFL-3}, where  $Z$ is plotted versus $\delta=1-n$ for various $U$. One sees that   Z decreases upon with increasing $U$, taking   a non-zero value  in  the $U=\infty$ limit. In this limit  its density dependence is  close to the empirical formula  $Z\sim \delta^{1.39}$. 

In the case of the 2-d $t$-$t'$-$J$ model the  ECFL results\cite{SM}  have a similar character.
The reduction of $Z $ from unity occurs as we approach the insulating limit $n\to 1$. Additionally, it  is very  sensitive to the sign and magnitude   of $t'/t$. The   dependence of $Z$ on $n$ and $t'/t$ is most clearly seen in Fig.~(1) of \cite{SM}. Qualitatively we find that  $Z$ decreases when $t'/t$ is negative and growing in magnitude, whereas a positive $t'/t$ enhances its value.  Within  the theory, reduction of the magnitude of $Z$ i.e.  the loss of weight of the quasiparticles, is  compensated exactly by the growth of the background pieces of  the spectral function, as seen in Figs. (1-2) of \cite{MS}. We note that experiments on cuprates strongly indicate the growth of background weight, and indeed the ECFL theoretical results closely match experiments in regard to the shapes of spectral functions \cite{ECFL-expts}. 

 }

 The   resistivity calculations in Refs.\cite{SM,MS} were performed for a typical set  of model parameters chosen for illustrative purposes. In these works we  noted that the resulting  resistivities  are     broadly comparable to experiments in their magnitude  and on the  scale of  temperature variation. In the present paper we  push this observation   to a more explicit and quantitative level, by comparing the ECFL results of \cite{SM,MS} with experiments on  a few representative  high T$_c$ materials with both hole and electron doping. 
Although broken symmetries  of various types are possible within the methodology,  we focus here on  the properties of the paramagnetic normal state.

\vspace{.15in}

{\bf \S Parameters of the model:}

\vspace{.15in}

The ECFL theory results used here \cite{SM,MS} are valid for a quasi two-dimensional correlated metal, with separation $c_0$ between layers. The resistivity in the calculations \cite{SM,MS} arises from intrinsic  inelastic e-e scattering with the {\em umklapp} processes, inherent in the tight binding model,  relaxing the momentum efficiently.
The (smaller)  $a$ and $b$ axis lattice constants cancel out in the formula for  resistivity. The theory   gives the planar  resistivity in the form 
\beq \rho = R_{vK} \times  c_0 \times \; \bar{\rho}(  \frac{t'}{t}, \frac{J}{t}, \frac{k_BT}{t}, \delta). \label{eq1} \eeq 
where $R_{vK}=\frac{ h}{e^2}=25813\,\Omega$, is the von-Klitzing resistance.
The (dimensionless) theoretical resistivity $\bar{\rho}$ is   a function of  the four displayed dimensionless variables. Detailed formulas leading to this expression can be found  in  Eqs.~(45,46) of \cite{SM}, and Eqs.~(12,13) of \cite{MS}. More precisely  $\delta$ is the concentration of holes measured from half filling i.e. $\delta=1-n$ and $n= \frac{N_e}{N_s}$ where $N_e$($N_s$) is the number of electrons (copper sites). At $\delta=0$ ($n=1$)  the model describes  a Mott-Hubbard  insulator. We discuss below the exchange parameter $J/t$, which plays a secondary role at the densities considered here.
 While three parameters $c_0,\delta,T$ are obtained from experiments directly,  ARPES constrains the parameter $t'/t$ from the shape of the Fermi surface in most cases. Given these, the remaining single parameter $t$ fixes the resistivity on an absolute scale. In addition a usually small and  $T$ independent (extrinsic) impurity resistance, usually arising from  scatterers located off the 2-d planes, must  be estimated separately.

In addition to $c_0$, the basic parameters of the model are  the nearest neighbor hopping $t$,  the second neighbor hopping $t'$, and   a superexchange energy $J$ within a a tight binding description of the copper d-like bands. The parameter $t'$ plays an important role in distinguishing between hole doped superconductors ($t'<0$) with a positive Hall constant   and the electron doped superconductors ($t'>0$) with a negative Hall constant. 
 The shape of the Fermi surface is sensitive to the ratio of the bare hopping parameters $t'/t$, if one assumes that interactions do not change its shape very much- this is largely borne out in ECFL theory.  For this reason ARPES can most often provide us with a good estimate for this parameter $t'/t$, although $t$ itself is not fixed by knowing the shape of the Fermi surface.
 We fix $J$ at a typical value  $0.17$t. At the densities we study here we find that the   magnitude of $J$  has a very limited influence on the calculated  resistivity, as  seen e.g.  in Fig.~(24) of \cite{MS}.    For the  single layer cuprate  systems,  one has  two  $Cu$-$O$ layer per unit cell  and therefore  the separation $c_0$  equals half     the c-axis lattice constant $c_L$ \cite{BL,Leggett}.    The applicability of the theoretical calculations  to systems with higher number of layers per unit cell, such as $Bi$-2212 or $YBa_2Cu_3O_{6-\delta}$    is less direct. It  requires making  further  assumptions relating   $c_0$ to the lattice constants. In order to avoid this  we confine ourselves to single layer systems.

The theoretical results tested here are found by ignoring  a possible superconducting or magnetic  state.   We have produced a grid of theoretical calculations for $t'/t=-0.4,-0.3,\ldots,0.3$ at several densities in the range $0.12 \leq \delta \leq 0.22$ surrounding the interesting regime of optimal doping $\delta\sim .15$. Since the theory is smooth in most theoretical parameters we can interpolate in it, when necessary. Calculations are carried out in a wide range of $T$ with a lower end $T$$\sim100$K with a system size of $62\times 62$. Lower T calculations require bigger system sizes which are computationally expensive and alternate methods are possible for estimating the resistivity.  For example at  lower   T $\lessim 50$K the resistivity can be extrapolated to a quadratic in $T$  quite accurately using $\rho= \alpha T^2/(1+ T/T_0)$ with suitable constants $\alpha,T_0$. This form is consistent with the $T\to0$ Fermi liquid character of the theory below  the (already low) $T_0$

\vspace{.15in}

{\bf \S The choice of systems:}

\vspace{.15in}

The lattice structure of the cuprates allows for a systematic change in carrier concentration by chemical substitution of elements situated  away from the copper oxide planes,   without  severely impacting the  impurity resistance. The role of block layers or charge reservoirs in hosting the donors away from the copper oxide planes, plays an important role in achieving this property of  the cuprates\cite{BL,Leggett}. This feature also provides a useful  handle in our analysis, 
 we can access  data on families of cuprates that contain  a reasonably large range of electron densities.  Since the basic parameters of the theoretical model can be assumed unchanged with doping \cite{lattice-constant-change}, such a family provides a systematic proving ground for theory.  Thus the experimental  data used for testing the theory is narrowed down to the available  systematic sets of resistivity data on single layer cuprates with varying densities.

In Table.~(\ref{table}) we list the single layer cuprate compounds where data sets with several densities are available.  The  hole doped LSCO and BSLCO materials are well studied by many authors,  and the   data set  from Ref.~(\cite{Ando})  used here, reports a very extensive set of densities for each family. This  provides us with 11 densities for LSCO in the range $0.12 \leq \delta \leq 0.22$, and 7 densities for BSLCO in the range $0.12 \leq \delta \leq 0.18 $
which are essentially within the range treatable by  theory. We include recent thin film data on the electron doped  LCCO from Ref.~(\cite{LCCO}). Here 4 densities are available in the theoretical range and the very regular $\rho\sim +T^2$ type behavior of the data allows for  easily eliminating the impurity contribution. For a more balanced representation of the electron doped materials, we included data on NCCO from Ref.~(\cite{NCCO}).
The NCCO family contains only two densities in the theoretically accessible range,  of which one is  impacted by 2-d localization effects. It is therefore not as constraining as the other families. The choice of the above four families of single layer cuprates, with $\gssim 20$ sample  densities seemed sufficiently representative for our task.

In addition to the above set of materials there are a few others belonging to the single layer class with data provided for several densities. Amongst these  we have excluded from our analysis, the mercury compound Hg1201 ($HgBa_2CuO_{4+\delta}$) \cite{Pelc}
and  the  thallium compound Tl2201  ($Tl_2Ba_2CuO_{6+\delta}$) \cite{Thallium,Kokalj}. In the literature for these compounds,  the value of $T_c$ for different  samples   is quoted and one needs to extract the electron density from other measurements, e.g. the Hall constant.
This was hard for the author to achieve, with a required accuracy in density $\Delta \delta\sim 0.1$ necessary for the present analysis.

 In Table.~(\ref{table}) we quote the c-axis lattice constant $c_L$ taken from x-ray diffraction data. The ratio $t'/t$ is taken from angle resolved photoemission (ARPES) experiments on the shape of the Fermi surface, fitted to a tight binding band. In some cases the experimental fits include a small further neighbor hopping as well, we neglect it here since the corrections only fine tune the shapes of the Fermi surface while preserving their basic topology.
Theoretical estimates from band structure \cite{Markiewicz,Pavarini} are roughly consistent with the above experimentally guided choices of $t'/t$. 
\vspace{.25in}

{\bf \S: Protocol for fixing $t$ and estimating the impurity resistivity:}

\vspace{.25in}

 We  determine the magnitude of $t$ for each material by collating a data set consisting of experimental $\rho_{exp}(T,\delta)$ points at various densities $\delta= \delta_1, \delta_2, \ldots $. From this set we extract the   slope of the resistivity 
 \beq
 \Gamma(T^\Phi)= \left( \frac{d \rho_{exp}(T,\delta=0.15)}{ dT} \right)_{T=T^\Phi}. \label{Gamma}
 \eeq
 Equating $\Gamma(T^\Phi)$ to the corresponding theoretical slope at  $T^\Phi$ determines the single parameter $t$.
 The density is chosen as $\delta=0.15$  since it is in a regime where the calculation  is quite reliable. 
  $T^\Phi$ is chosen as  the midpoint of the temperature range of the data set,  so that  $T^\Phi= 250$K for LSCO and $T^\Phi=200$K for BSLCO and LCCO in the following analysis.

 We next  need to estimate the $T$ independent impurity contribution to the resistivity at each density $\rho_{imp}(\delta)$ for LSCO and BSLCO \cite{comment1}. For LCCO the impurity contribution $\rho_{imp}$ has been eliminated by the authors of \cite{LCCO} thereby this task is already done. For the others we shift down the experimental resistivity $\rho_{exp}(T^\Phi,\delta)$ to match the theoretical resistivity, the magnitude of the shift  gives us the estimated $\rho_{imp}(\delta)$ at each density. We are thus using the relation $\rho_{exp}(T^\Phi,\delta) - \rho_{imp}(\delta)=\rho_{th}(T^\Phi,\delta)$, where $\rho_{th}$ is from \disp{eq1}.  The impurity contribution is displayed in all figures and is a small fraction of the total resistivity in all cases.

In summary  fixing the magnitude of $t$ for a data set  requires a comparison with experiments at a {\em single density} ($\delta=0.15$)  and  a {\em single temperature} ($T=T^\Phi$). The impurity contribution is estimated at each density at the same temperature $T=T^\Phi$. Checking these against data constitutes the essence of the test carried out here.  The final two columns  in Table.~(\ref{table}) report the  fitted value of the single undetermined  parameter $t$. The  bare band width  is estimated as  W$\sim$8 t. Slightly different choices of the density and $T^\Phi$ lead to comparable results for $t$.

Before looking at the results, we make a few comments about the analysis. (a)
 The  requirement that the fitted values of $t$ and $t'/t$ remain   unchanged for  different densities $\delta$ gives added significance to the fits. It is clearly 
  an important and non-trivial requirement from any theory as well.  In this sense  matching  the experimental resistivity at a single density of any particular compound,  is  less significant than doing so at a sequence of different densities. (b) The impurity shifts  reported in each curve, are seen to be on a typically   expected scale  $\sim$50-150$\mu\Omega$cm. The data on LCCO \cite{LCCO} is available with the impurity contribution already removed by the authors. 
  (c) At low electron densities the effects of (2-d) electron localization are visible in some data sets. In these cases the  impurity contribution leads to an upturn at low T. This upturn has been discussed extensively in literature \cite{Ando,NCCO} and also manipulated with magnetic fields \cite{Boebinger-LSCO}.  Since the ECFL  theory excludes any strong disorder effects, we do not expect to capture these in the fits.

 \begin{table}
\centering
\begin{tabular}{||p{3.2cm} |p{1.3cm}|p{1.3cm}|p{1cm}||}
\multicolumn{4}{c}{{\bf  Single Layer High $T_c$ Compounds}}  \\ \hline \hline
{  Material} & $c_L$\; $A^0$ & $t'/t$  & t (eV) \\ \hline \hline
 $La_{2-x}Sr_xCuO_4$ (LSCO)&13.25 \cite{LSCO-c,Ando} & -0.2 \; \cite{Yoshida-LSCO,Ando-Hashimoto} & 0.9  \\ \hline
 $Bi_{2}Sr_{2-x}La_xCuO_6$ (BSLCO)&24.3 \cite{Ando} & -0.25 \cite{Ando-Hashimoto} & 1.35 \\ \hline
 $Nd_{2-x}Ce_xCuO_4$ (NCCO)&12.01 \cite{BL,NCCO-c} & +0.2 \; \; \cite{DMKing} & 0.9  \\ \hline
 $La_{2-x}Ce_xCuO_4$ (LCCO)&  12.45 \cite{Sawa}& +0.2 & 0.76  \\ \hline
 \hline 
\end{tabular}
 \caption{ \footnotesize The single layer cuprates analyzed in this work. For the first three materials the values of $t'/t$ are obtained from ARPES experiments where the Fermi surface shape is fit to a tight binding model. 
  Here LCCO data is from thin films while the others are from single crystals.  $c_L$ is the c-axis lattice constant. In all the above cases the unit cell contains two copper oxide layers, and hence  their separation  $c_0$  entering  \disp{eq1} is half the lattice constant $\frac{1}{2} c_L$ \label{table}.  For LCCO the  ARPES data on the Fermi surface  does not exist. The quoted    $t'/t$  is chosen to be the same as NCCO.    The last column lists the values of  $t$ determined in this work.  The single adjustable parameter the hopping  $t$ is  found using the slope of the experimental resistivity at 200K at a single density $\delta=0.15$.  Band structure \cite{Markiewicz,Pavarini} estimates of $t'/t$  ratio are quite close to the ones used here, but the estimates of $t$  differ somewwidetilde.   It must be kept in mind  that the quoted parameter $t$ is the  bare one, i.e prior to   many-body renormalization. }
\end{table}


For LCCO the digital data was kindly provided by  the authors of \cite{LCCO}. For the other data sets studied here  
 the published resistivity data was  digitized  using the commercial software program {\it DigitizeIt} \cite{Borman}. We found that the program works quite well provided the experimental curves do not overlap or cross. This feature limited our data extraction to some extent, as the reader might notice from the low temperature truncation in the experimental data in the figures presented below.

  \begin{figure}[htbp]
\centering
\subfigure[LSCO-Slightly underdoped to optimally doped]{\includegraphics[width=\columnwidth]{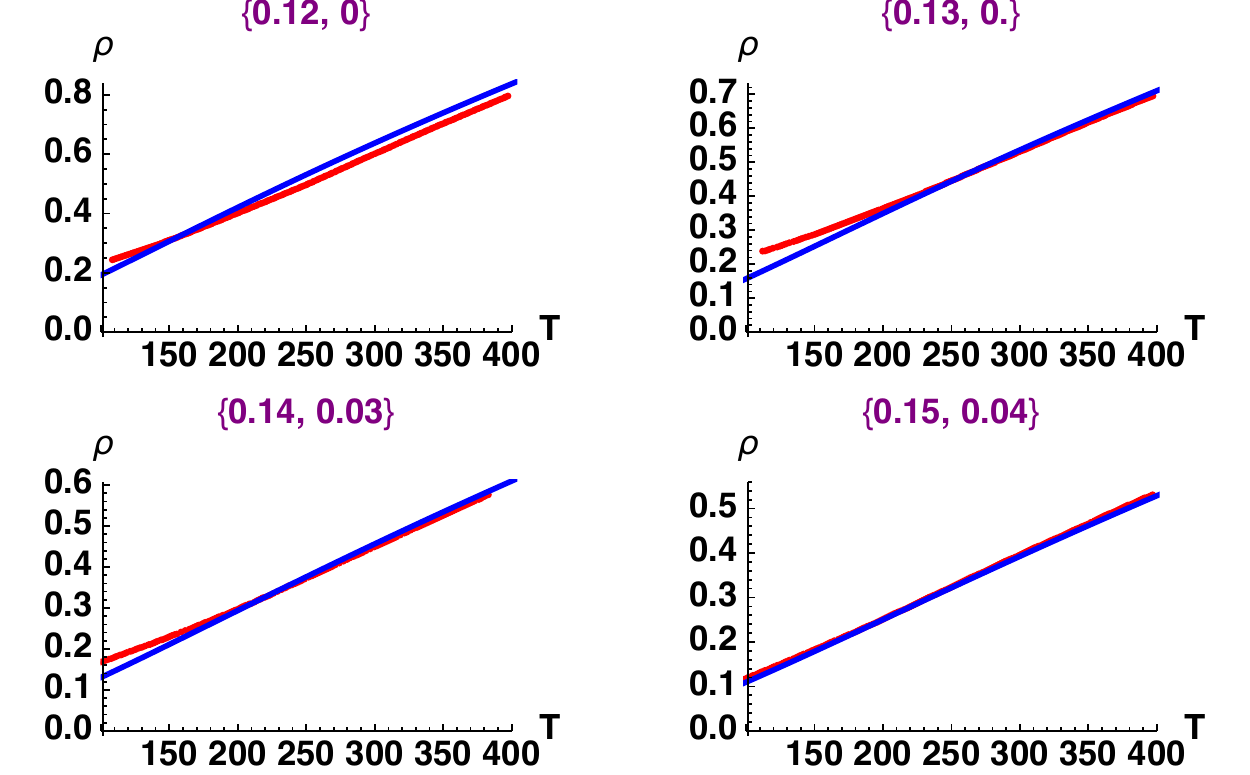}}
\subfigure[LSCO-Near Optimal Doping]{\includegraphics[width=\columnwidth]{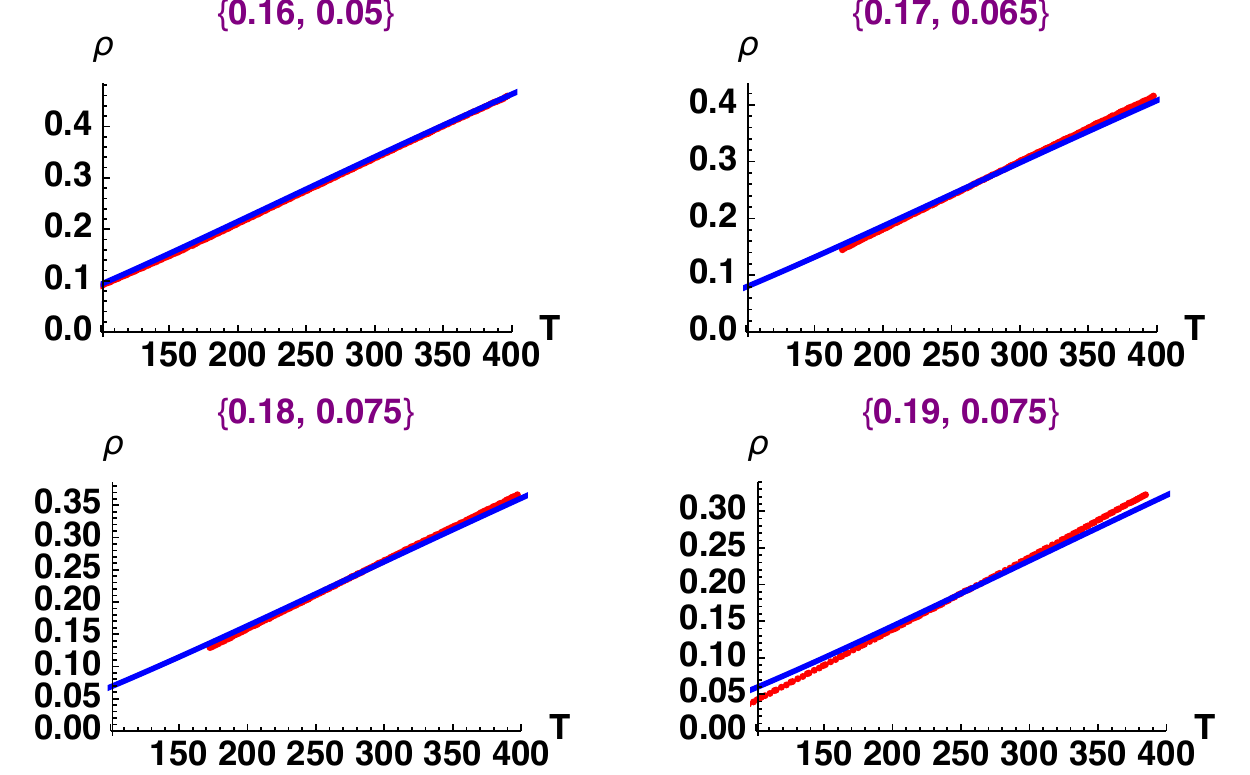}}
 \caption{\footnotesize
 Dotted (red) line is data extracted  from Y. Ando et.al. \cite{Ando} Fig.~2(b), and solid (blue) curve is the theoretical curve with t'/t=-0.2.   The displayed  pair of numbers  $\{ \delta, \rho_{imp}\}$  indicates the hole density   and  estimated impurity resistivity.   
 The resistivity $\rho$ is everywhere  in units of  m$\Omega$cm and T is in Kelvin. The parameter $t=0.9$ eV  was fixed using $\Gamma(T^\Phi)$  the slope of the  resistivity at $\delta=0.15$, $T^\Phi=250$ K (see \disp{Gamma}).  The resistivity at every   density in this plot and in \figdisp{Fig-LSCO-2} is  then predicted by the theory. 
 \label{Fig-LSCO-1} }
 \end{figure}
 \begin{figure}[htbp]
\centering
\subfigure[LSCO-Slightly overdoped]{\includegraphics[width=1\columnwidth]{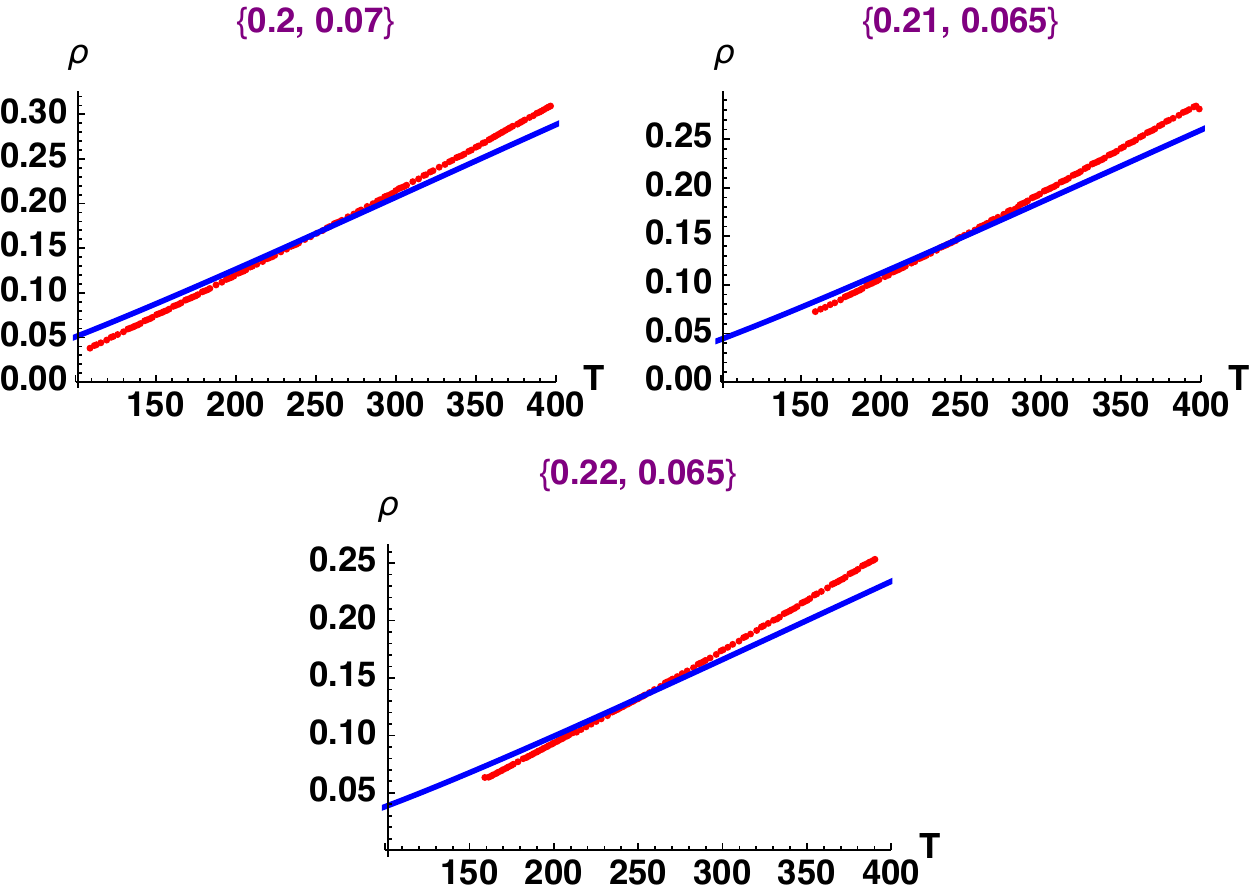}}
 \caption{\footnotesize
(a)  Dotted (red) line is data extracted  from Y. Ando et.al. \cite{Ando} Fig.~2(b) for the slightly overdoped cases, and solid (blue) curve is the theoretical curve with $t'/t=-0.2$ and $t=0.9$ eV.    \label{Fig-LSCO-2} }
 \end{figure}


We next describe the comparison for different systems. 

\vspace{.15 in}

{\bf \S LSCO:}

\vspace{.15 in}

In \figdisp{Fig-LSCO-1} (a,b) and \figdisp{Fig-LSCO-2} (a) the extensive dataset from \cite{Ando} FIg.~2(b) is compared with theoretical predictions. The parameters in Table~(1) are used here.
The band parameter t= 0.9 eV is found from the slope of the resistivity  $\delta=0.15$, $T^\Phi$=250K. All other densities are then predicted by theory on an absolute scale.  While some deviations at low density $\delta=.12$ and also at high density $\delta\gssim 0.2$ are visible, the overall agreement seems fair.    For the same parameters \figdisp{Fig-Theoretical} (a) shows the theoretical resistivity over an enlarged temperature window. Here subtle changes of curvature are visible at high and low T.

\vspace{.15 in}

 { \S { \bf BSLCO} and {\bf $Bi$-$2201$}:}
 
\vspace{.15 in}

\addtolength{\voffset}{-1.5cm}
 \begin{figure}[htbp]
\centering
\subfigure[BSLCO-Slightly underdoped]{\includegraphics[width=\columnwidth]{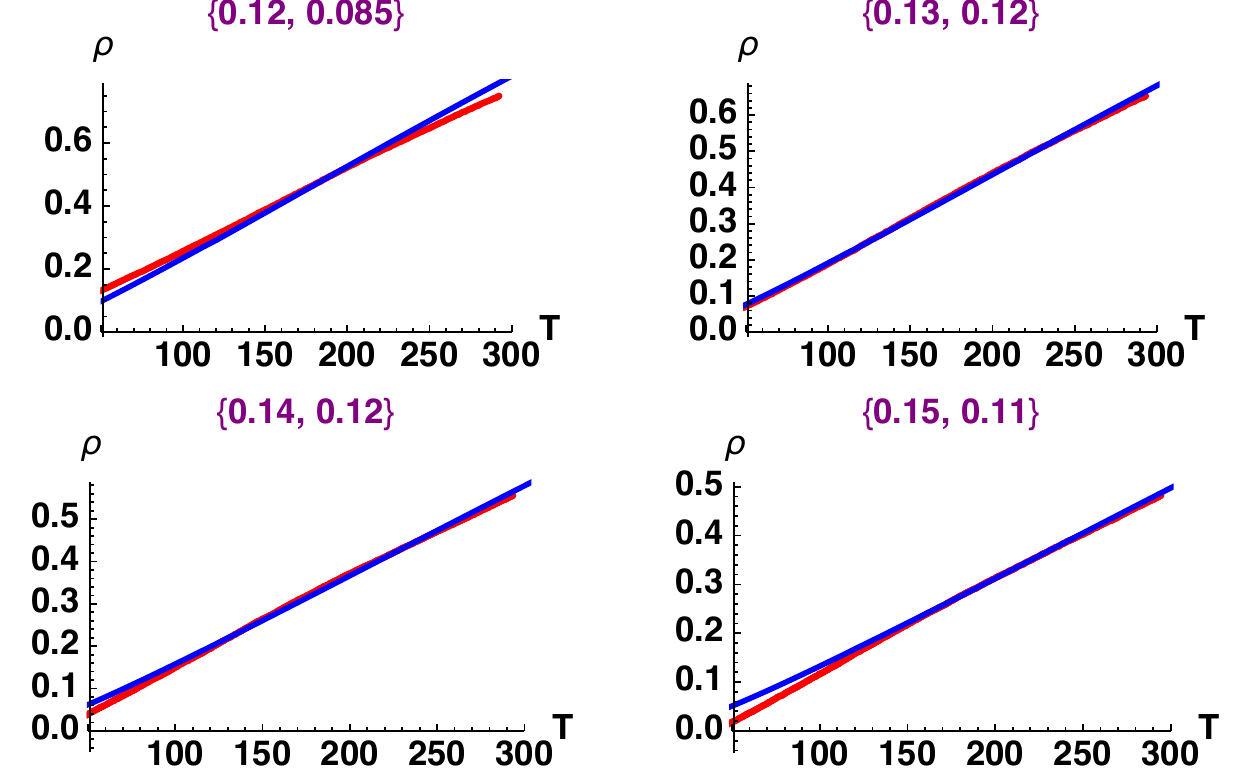}}
\subfigure[ BSLCO- Near Optimal doping ]{\includegraphics[width=\columnwidth]{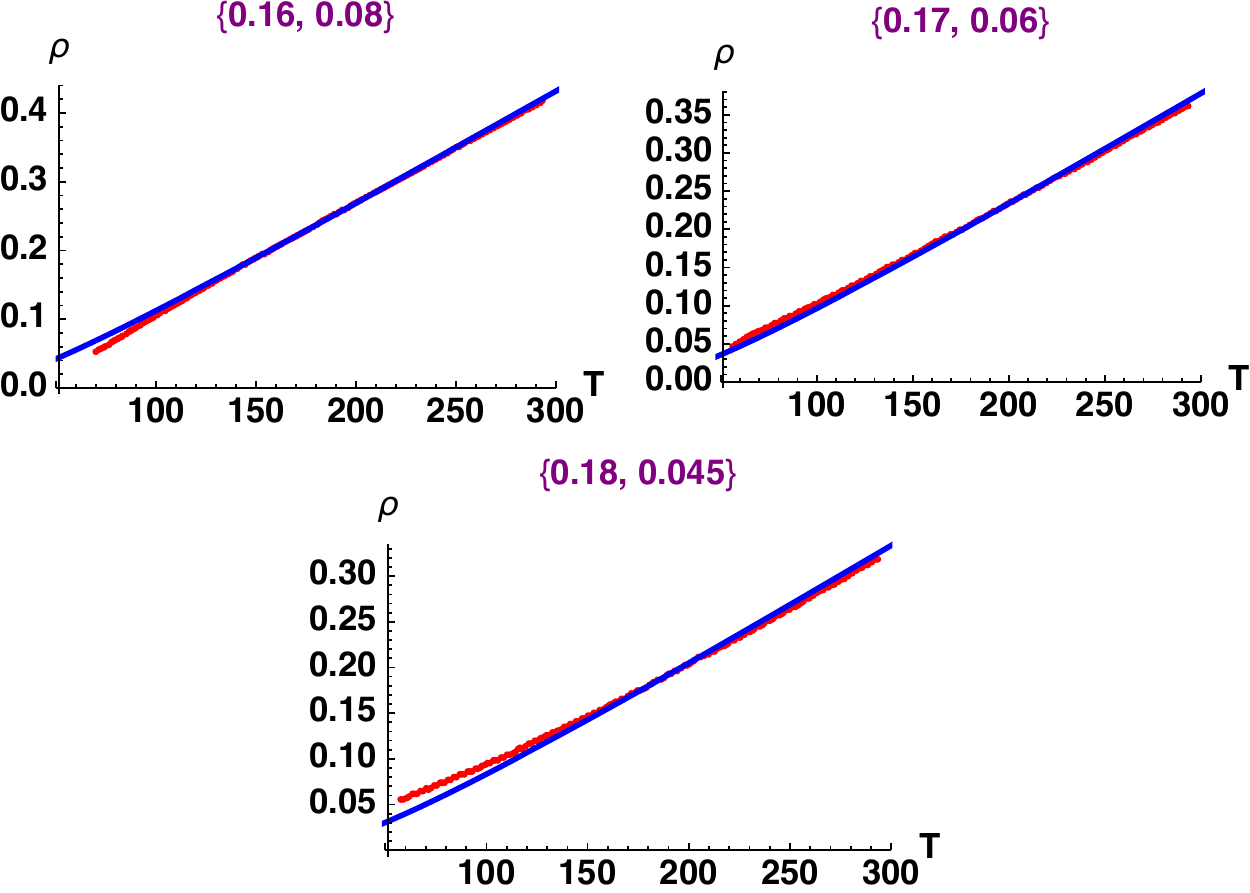}}
 \caption{\footnotesize  Dotted (red) line is data extracted  from Y. Ando et.al. \cite{Ando} Fig.~2(b), and solid (blue) curve is the theoretical curve with $t'/t=-0.25$.  The displayed  pair of numbers  $\{ \delta, \rho_{imp}\}$   indicates the hole density   and  estimated impurity resistivity.   
 The resistivity $\rho$ is   in units of  m$\Omega$cm and T  in Kelvin. The parameter $t=1.35$ eV was fixed  using  using $\Gamma(T^\Phi)$  the slope of the  resistivity at $\delta=0.15$, $T^\Phi=200$ K (see \disp{Gamma}). The resistivity at every  density is then predicted by the theory. 
 \label{Fig-BSLCO-1} }
 \end{figure}
\addtolength{\voffset}{1.5cm}


In \figdisp{Fig-BSLCO-1}  the data for  the BSLCO family  of compounds  $Bi_2Sr_{2-x}La_xCuO_6$ from Y. Ando Ref.~(\cite{Ando}) is compared with theory. The band parameters t=1.35 eV is  found from  the slope of the resistivity at $\delta$=0.15, $T^\Phi$=200K. All  other densities are then predicted by theory. For these  parameters \figdisp{Fig-Theoretical} (b)  shows the theoretical resistivity over an enlarged temperature window. The  larger value of $t$ in BSLCO relative to that in LSCO can be understood from comparing \figdisp{Fig-Theoretical} (a) and (b). The almost doubled value of $c_0$ increases by a similar factor  the resistance of BSLCO over that of LSCO, provided one is at  the same scaled temperature $T/t$. A larger $t$   spreads this increase over a larger T window.

\vspace{.15 in}

{ \bf \S NCCO and LCCO:}

\vspace{.15 in}

The NCCO family of materials with composition
 $Nd_{2-x}Ce_xCuO_4$ and the closely related LCCO family $La_{2-x}Ce_xCuO_4$
  are of considerable interest as  counterpoints to the other two families studied above. Both have  the opposite sign of the Hall constant  from the hole doped cases and display a pronounced $T^2$ type resistivity.

  In a single band model description, such as the $t$-$t'$-$J$ model used here, these materials can \change{also}  be treated as having a filling \change{less} than half. \change{ The filling of these materials in the original electron picture is greater than half. Starting with a Hubbard  model   one can perform a  particle hole transformation of both spin species  to map the model   to less than half filling.  For U large enough  the \tJ model is  once again introduced in the place of the Hubbard model. This process generates some U dependent constant terms that are absorbed into the chemical potential. It also flips the sign of all hopping matrix elements. While the nearest neighbor hopping $t$ can be flipped back to the standard (positive) sign using a simple unitary transformation  (exploiting the square lattice geometry), the second neighbor hopping $t'$ is now positive and the Fermi surface is  electron-like.  }
  
\begin{figure}[htbp]
\centering
\includegraphics[width=.9\columnwidth]{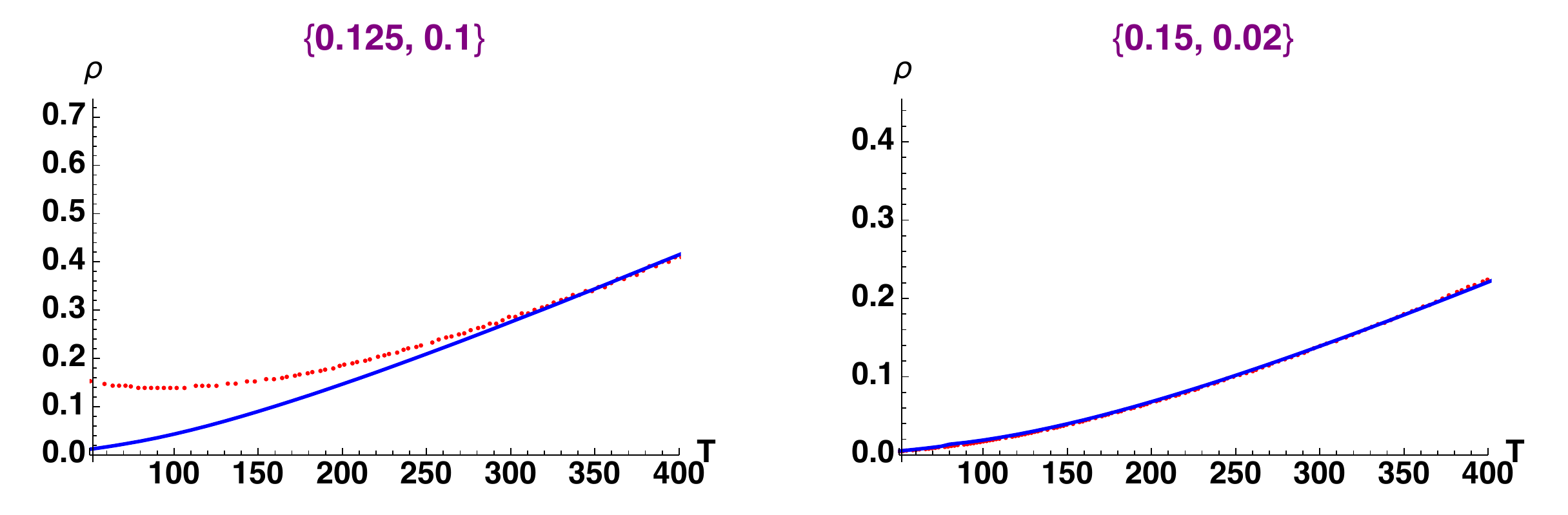}
 \caption{\footnotesize  Dotted (red) line is data extracted  from Y. Onose {\it et.al.}  \cite{NCCO} Fig.~9(b), and solid (blue) curve is the theoretical curve with $t'/t=+0.2$.  The parameter $t=0.9$ eV was fixed using the  slope of  the $\delta=0.15$ data at 200K.
The data set contains only these two densities within the range accessible to theory. The upturn in the lower density curve and the larger magnitude of impurity resistivity is due to strong disorder effects, as already noted in \cite{LCCO}.  The sign of $t'/t$ is reversed between this   figure  and \figdisp{Fig-LSCO-1}  for LSCO, while  other parameters $c_0, t$ are essentially unchanged.   Both the experimental and  theoretical resistance display a resistance with a positive upward curvature ( i.e.  $\rho \sim+T^2$). 
 \label{Fig-NCCO} }
 \end{figure}

  On the materials side, the available data on NCCO \cite{NCCO} (see Fig.~(9b))  is relatively sparse in the metallic range containing only two  samples. One of these   is afflicted with strong disorder effects at  low T. In \figdisp{Fig-NCCO} we compare the data from Y. Onose {\em et. al.} \cite{NCCO} with theory. While the density $\delta=.15$ is perfectly matched with theory, the lower density $\delta=.125$ curve shows a distinct upturn at low T, as  discussed in \cite{NCCO}.  A systematic treatment of strong disorder effects in the ECFL theory is currently missing.

\begin{figure}[htbp]
\centering
\includegraphics[width=\columnwidth]{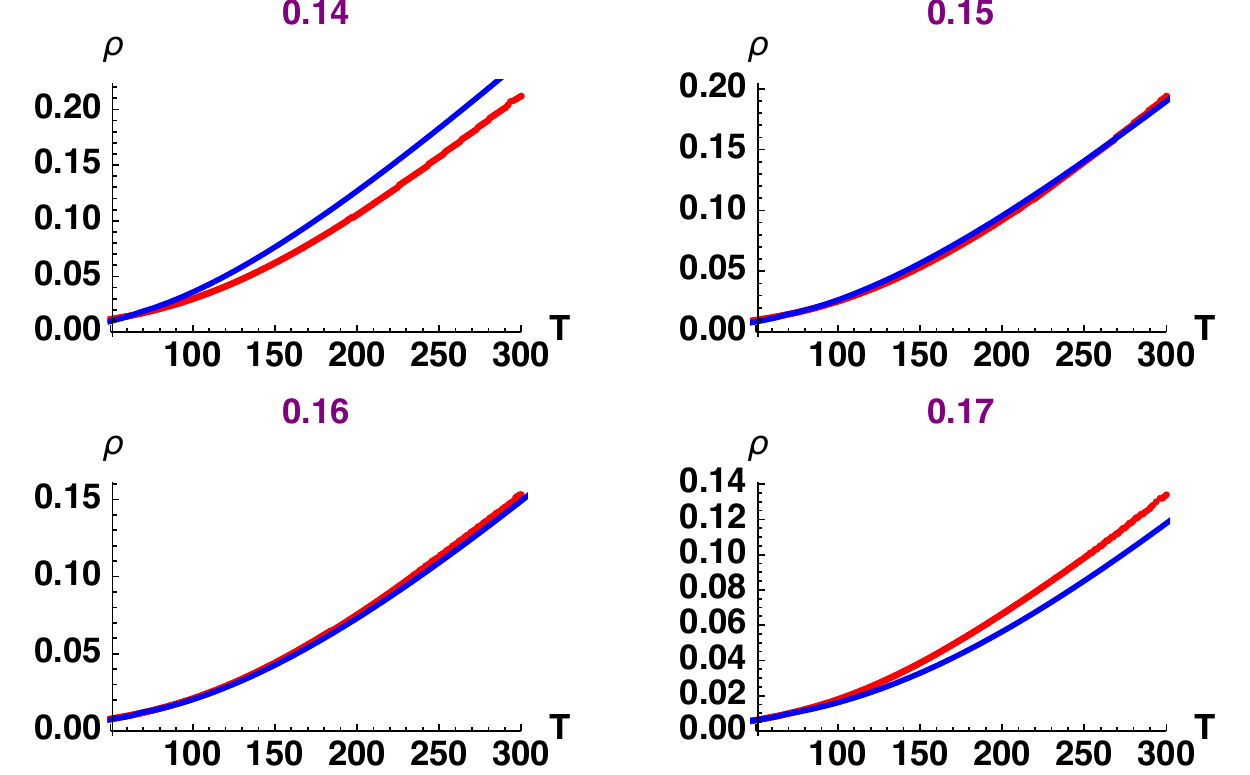}
 \caption{\footnotesize  Data  is from T Sarkar {\it et.al.}  \cite{LCCO} Fig.~2(b)
  as the  dotted red line. The impurity contribution in this data set has been removed by the authors in \cite{LCCO}.
  The theoretical curve is in solid blue,  with $t'/t=+0.2$.  The hole density is marked at the top in each plot.
 The parameter $t=0.76$ eV was fixed using using $\Gamma(T^\Phi)$  the slope of the  resistivity at $\delta=0.15$, $T^\Phi=200$ K (see \disp{Gamma}).
 The sign of $t'/t>0$ is  common to  NCCO  and  reversed from that in   LSCO and BSLCO. Both  experiments and  theory find a resistance with a positive curvature ( i.e.  $\rho \sim+T^2$), as in NCCO. This is in striking contrast to LSCO and BSLCO as seen in Figs.~(\ref{Fig-LSCO-1},\ref{Fig-LSCO-2}) and Figs.~(\ref{Fig-BSLCO-1}). 
 \label{Fig-LCCO} }
 \end{figure}

\begin{figure}[htbp]
\centering
\subfigure[LSCO:   $\delta=$0.12$\to$0.22($\downarrow$)]
{\includegraphics[width=.49\columnwidth]{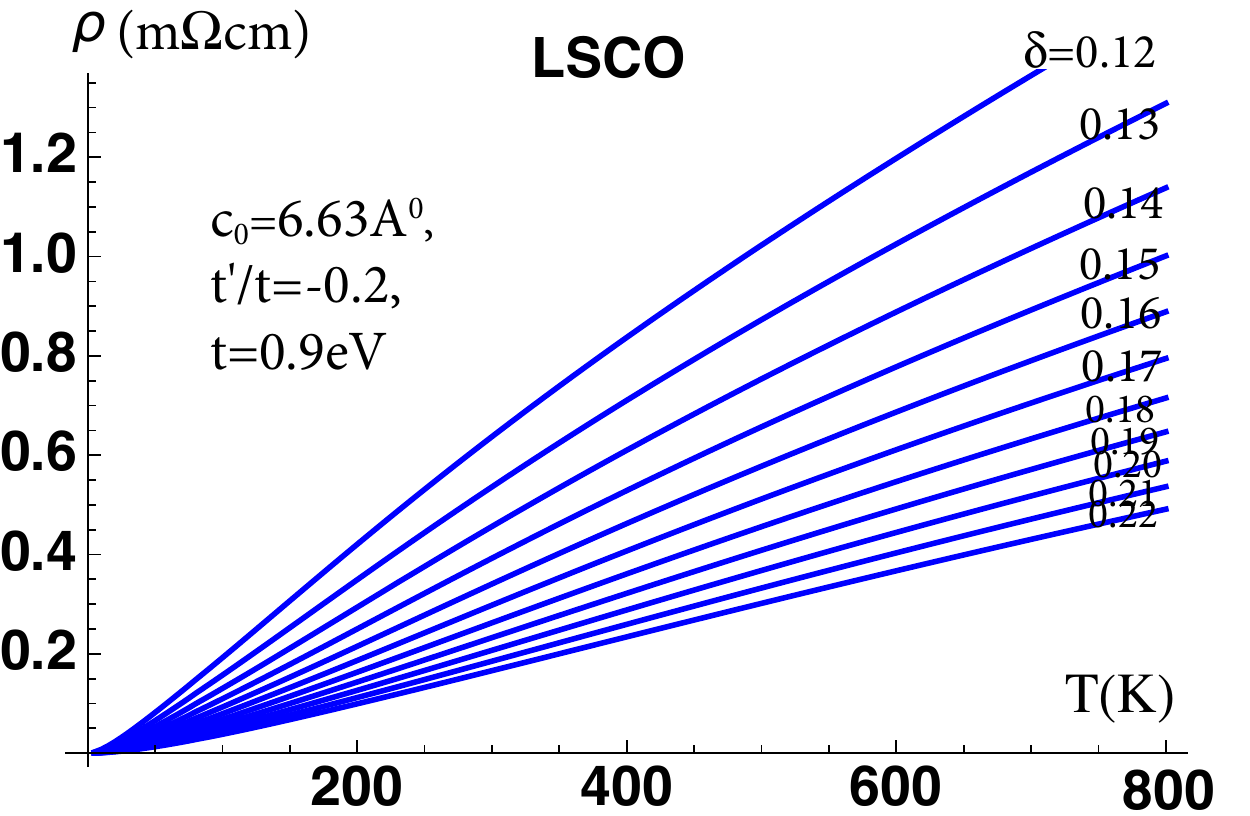}}
\subfigure[BSLCO:   $\delta=$0.12$\to$0.22($\downarrow$)]
{\includegraphics[width=.49\columnwidth]{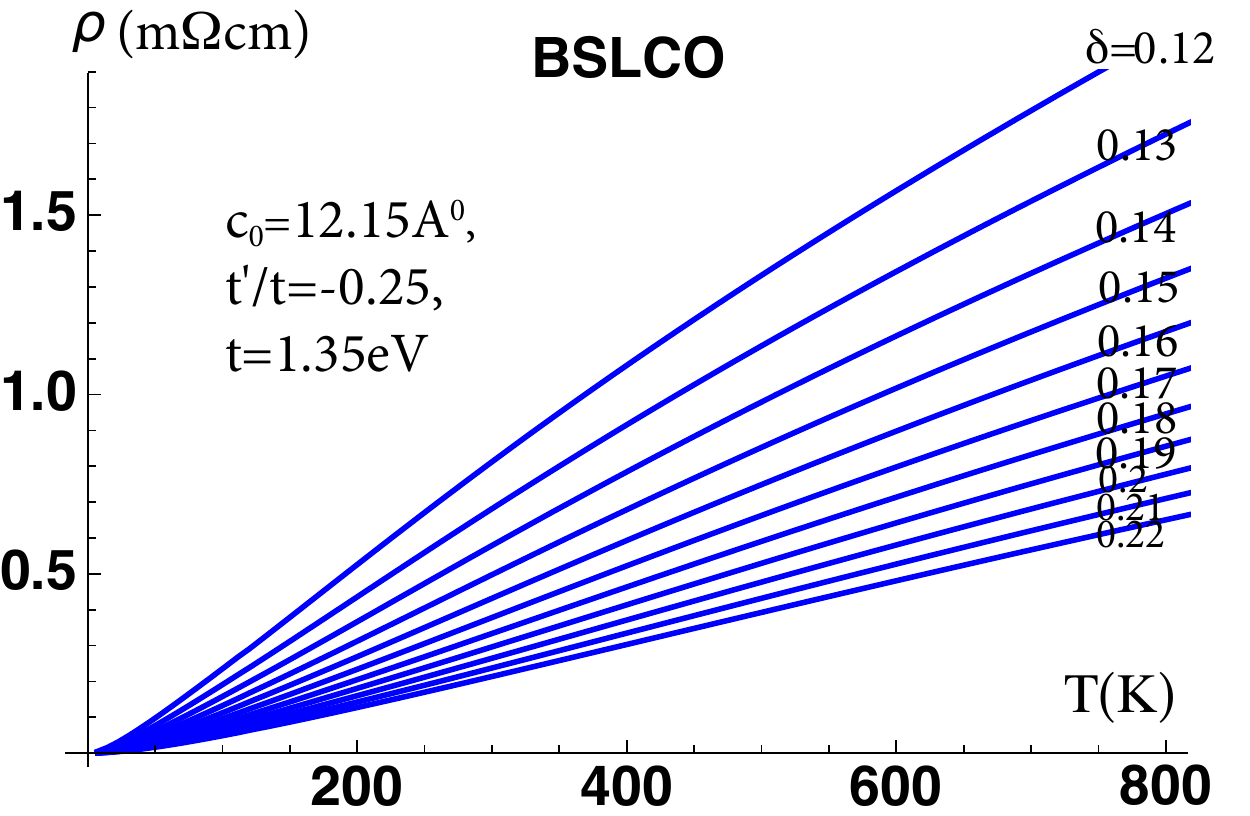}}
\subfigure[LCCO:  $\delta=$0.12$\to$0.2($\downarrow$)]
{\includegraphics[width=.49\columnwidth]{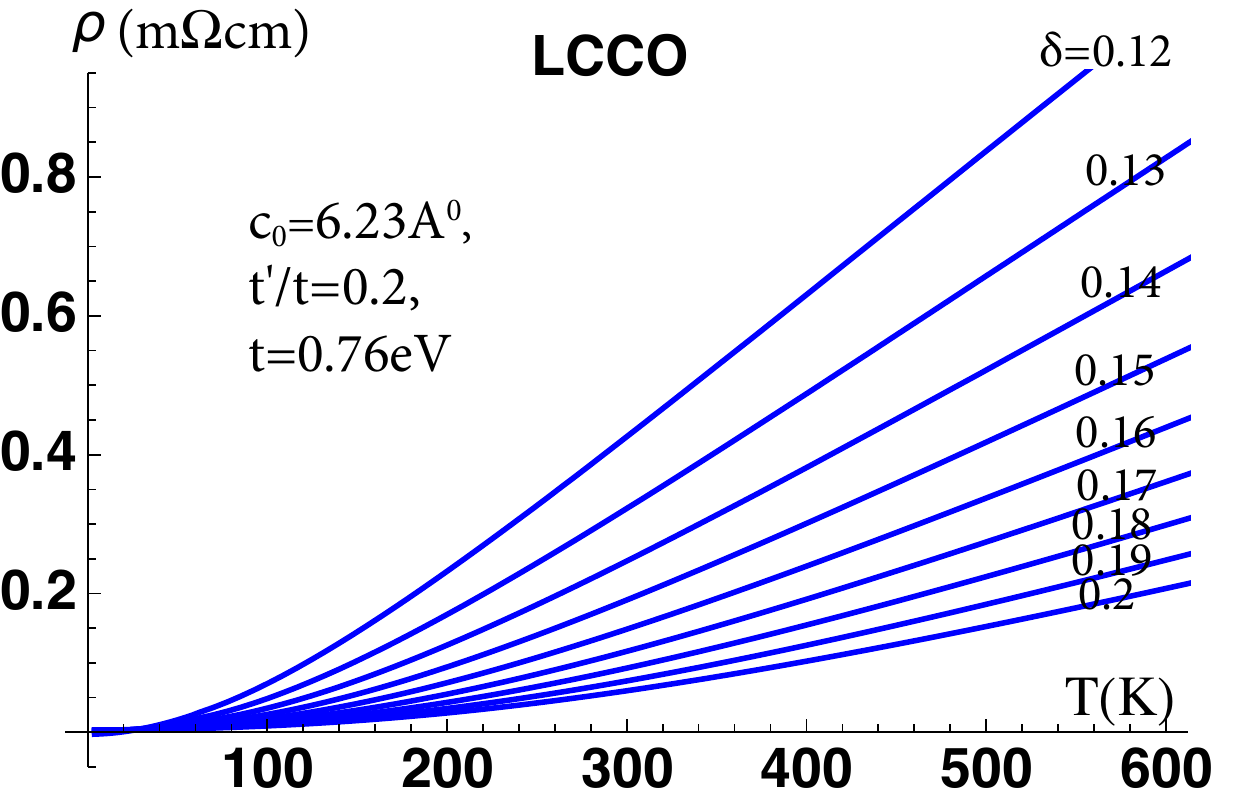}}
 \caption{\footnotesize
Theoretical resistivity curves  for LSCO,  BSLCO and LCCO  over an extended temperature range. In going from LSCO with BSLCO   $c_0$,  i.e.  the  separation between  the layers, is  almost doubled while $t'/t$  changes only slightly. The resistivity at a comparable $(\delta,T)$
here, and also in the data, changes by a smaller factor than $c_0$. In order to reconcile with  this feature of the data,  the deduced hopping parameter $t$  is greater  by $\sim$50\%  for BSLCO relative to LSCO. The distinct T dependence of LCCO relative to the other two systems is striking. Additionally it is  noteworthy that the intrinsic resistivity of the electron doped  LCCO is considerably smaller than that of the hole doped LSCO. Since these have roughly the same $c_0,t, |t'/t|$ values, the difference is attributable to the different  sign of $t'/t$.
   \label{Fig-Theoretical} }
 \end{figure}

  The data on  LCCO\cite{LCCO}  gives us four densities within the range covered by theory. In the absence of ARPES data we chose $t'/t=0.2$, i.e. the same value as in NCCO.  We have verified that nearby values to $t'/t$ lead to a similar quality of fits after adjusting the parameter $t$, and hence this choice not final. The authors conveniently present the  resistivity in Fig.~ 2(b) of \cite{LCCO} requiring no further impurity corrections. In \figdisp{Fig-LCCO} we compare theory and experiment, and in \figdisp{Fig-Theoretical} (c) we present the theoretical resistivity on an extended T scale at several  densities.  The  discrepancy in LCCO between theory and experiment at $\delta=0.17$ at  T=200 is $\sim 0.01$, and is quite visible. However  we should keep in mind that at corresponding densities the absolute scale of the resistivity for LCCO is considerably smaller than that for  LSCO and BSLCO. This can be seen  in  \figdisp{Fig-LSCO-2} and \figdisp{Fig-BSLCO-1}. As a consequence a similar scale of absolute error       leads to a much larger relative error.

\vspace{.15 in}

 {\bf \S Discussion:}
 
 \vspace{.15 in}
 
 
 We have presented a comparison of theoretical resistivity  with extensive data on three  families of cuprate superconductors.  It is also feasible to fit data on non-cuprate strongly correlated systems such as 
   $Sr_2RuO_4$ from \cite{SRO}, where data over  a large range $ T \leq 1000$  is available. However data is available at only one composition in this case, and the value of $t'/t$ is hard find from experiments. Since a single  density within a family  does not test the theory  stringently, we  omit the comparison here. 

Overall we have shown that the ECFL theory gives a  reasonable account of data  in the three families discussed above. A small number of parameters taken from experimental data  fix the  model completely.  It is encouraging that 
  the resulting resistivity  affords a reasonable  fit to a collection of resistivity  data  at various densities, both in terms of the T dependence and its magnitude. It is also encouraging  that upon using different model parameters, the same calculation  fits the resistivity of both hole doped and electron doped materials.

In \figdisp{Fig-Theoretical} we
display  the theoretical  resistivities on a larger  T scale and for more densities, using parameters of the three families separately. We found that the data is fit almost equally well by making nearby choices of the pair $t'/t$ and $t$.  The differences between different choices do exist and  show up but only at higher T, especially in the location of subtle kinks of the sort seen in \figdisp{Fig-Theoretical}.

\vspace{.15in}

{\bf \S: Conclusions:}

\vspace{.15in}

From the above exercise it appears that the extremely correlated Fermi liquid theory has the necessary ingredients to explain the variety of data seen in the above materials. Other materials, some of them with higher number of layers, do display further subtle features which are missing in the theory. However these features are also missing  in  the displayed data from the above  materials. We have thus made a fair beginning with the above  ``standard'' cuprate materials, but further challenges from more complex behavior are to be expected.

 A few comments on the results and their implications are appropriate.  
  Let us first discuss the holed doped materials. Here  the quasi-linear resistivity seen near $\delta$$\sim$0.15 is remarkable, as noted by many authors. We should also pay attention to the underlying   suppression of scale. By this   we refer to the fact  that the  temperature scale of resistivity  variation is as low as $\sim$100-300 K, starting from a  bare band width of almost 10eV. The three orders of magnitude reduction in scale is non-trivial,  reminiscent of the  emergence of the low energy Kondo scale  in magnetic impurity systems.   Starting from wide energy  bands with a width of $\sim$ 10eV, the ECFL theory systematically generates low energy  and temperature scales, a few orders of  magnitudes smaller than the bare ones\cite{ECFL,ECFL-2,ECFL-expts}. The low energy scales
     depend sensitively  on the density and a few other  parameters, especially the sign and magnitude of $t'/t$.   
     
   A major part of this scale suppression is due to the  small quasi-particle weight $Z\lessim$$0.1$ at relevant densities that arise in the theory \cite{ECFL,SM,MS}. More physically we can attribute this suppression to the profound role of Gutzwiller projection  on the electron propagators near the Mott-Hubbard  half filled limit. It is    captured to a good extent  by the ECFL theory, and is visible in the   detailed structure of the electron spectral functions \cite{ECFL,MS,SM}

  For the electron doped materials, it is interesting that the theoretical resistivity matches  experiments essentially as well  as for the hole doped materials. The two classes of materials have   the opposite sign of the   parameter $t'/t$, which is disconnected from the extent of  correlations. This finding has a bearing on the frequently debated topic of  the Fermi liquid nature of electron doped cuprates. The ECFL theory says that both hole doped and electron doped systems are  (extremely correlated) Fermi liquids at the lowest temperature. Additionally
    the  theory quantifies   the range of T where a Fermi liquid type behavior $\rho\sim T^2$ holds good. Going further it also  identifies regimes  succeeding the Fermi liquid \cite{Strange-Metal,ECFL,ECFL-2,ECFL-Infinite-D,SM,MS} upon warming. 
    
    In order to better understand  the origin of the difference between hole and electron doping within the theory, the following observation may be helpful.   It is known    that  the sign and magnitude of the parameter $t'/t$ directly
 influences the magnitude of the already small quasiparticle weight $Z$ (see Fig.~(1) of \cite{SM}).  A positive $t'/t$ leads to a small   $Z$, while a negative $t'/t$ leads to an even smaller but non-vanishing $Z$.   For the electron doped case this distinction ultimately  results  in an enhanced thermal range displaying a positive curvature of the $\rho$-$T$ plots. The effect on resistivity of the sign of $t'/t$ can be seen explicitly by comparing 
the theoretical resistivity curves  for the hole doped cases \figdisp{Fig-Theoretical}(a) or (b)  with the electron doped case in \figdisp{Fig-Theoretical}(c).

As a crosscheck on the theory, it should be interesting to compare other physical variables  with data for the systems considered here, using the  deduced parameters.
Finally we  should note that future technical developments in the implementation of the ECFL theory are likely to refine some of the theoretical results presented here.

\vspace{.25 in}

 {\bf \S Acknowledgements:} 
  
  \vspace{.25 in}

We are grateful to Professor  A. J.  Leggett for   stimulating   discussions and for valuable comments on the manuscript. We  thank   Professor R. L. Greene and Dr.  T. Sarkar    for providing  the digitized resistivity data of \cite{LCCO}. We also thank  Professor M. Greven and Professor A. Ramirez for useful comments. 

The work at UCSC was supported by the US Department of Energy (DOE), Office of Science, Basic Energy Sciences (BES), under Award No. DE-FG02-06ER46319. The computation was done on the comet in XSEDE\cite{xsede} (TG-DMR170044) supported by National Science Foundation grant number ACI-1053575.


\begin{thebibliography}{99}
\bibitem{comment1} 
 Extrapolating the normal state resistivity to $T\to0$ is possible for some densities  where the $T$ dependence is purely linear (or purely quadratic) over most of the  range. In such a case extrapolation gives  the same result as our procedure. For general cases with a more complex $T$ dependence, extrapolation requires an indeterminate fitting function. Our procedure then corresponds to choosing the fitting function as the  set of figures in \figdisp{Fig-Theoretical}  (all vanishing as $T\to0$) plus an impurity contribution.
\bibitem{spectral-function-SM} The quasiparticle weight $Z$ is given at various fillings and $t'/t$ in Fig.~(1) of \cite{SM}. The rapid plunge of the spectral function peaks in the Brillouin zone with warming are shown in Fig.~(3) of \cite{SM}.  Note that $T$ is scaled by $t$, and   the value of $t$ in this figure is fixed at t=.45 eV in \cite{SM}. Therefore $T$ (and $t$) must be rescaled for  different materials as per the values  discussed in Table.~(1)
\end{thebibliography}
 \end{document}